\documentclass[12pt]{article}
\usepackage{amsmath}
\usepackage{amsthm}
\usepackage{amssymb}
\usepackage{latexsym}
\usepackage{amsfonts}
\usepackage{amsbsy}
\newcommand{\nc}{\newcommand*}
\newcommand{\rnc}{\renewcommand*}
\nc{\reff}[1]{(\ref{#1})}
\nc{\lan}{\langle} \nc{\ran}{\rangle} \nc{\ts}{\textstyle}
\nc{\ds}{\displaystyle}
\rnc{\Bbb}{\mathbb} \rnc{\rm}{\text}
\nc{\eps}{\epsilon} \nc{\vep}{\varepsilon} \nc{\vp}{\varpi}
\nc{\vpi}{\varphi} \nc{\vt}{\vartheta} \nc{\vr}{\varrho}
\nc{\wh}{\widehat} \nc{\wt}{\widetilde}
\nc{\hf}{{\mathfrak H}_F}
\def\cD{{\mathcal D}}
\def\cH{{\cal H}}
\def\cN{{\mathcal N}}
\def\chm{\cH_{\mu}}
\def\BR{{\Bbb R}}
\def\BC{{\Bbb C}}
\def\one{1\hskip-.37em 1}

\def\half{{\frac{1}{2}}}
\def\bra#1{\lan {#1} |}                           
\def\ket#1{|{#1} \ran}                            
\def\pair#1#2{\langle {#1} \mid {#2} \rangle} 
\def\qN#1{{\left[#1\right]}_q}      
\def\gex#1{\wh{e}_q(#1)}

\def\apm{a_{\mu}^+}
\def\amm{a_{\mu}^-}
\nc{\Szi}[1]{\sum_{ #1 = 0 }^{\infty}} \nc{\Sii}[1]{\sum_{ #1 =
-\infty }^{\infty}} \nc{\Szx}[2]{\sum_{#1 =0}^{#2}}
\nc{\Syx}[3]{\sum_{ #1 = #2 }^{#3}} \nc{\Pzi}[1]{\prod_{ #1 = 0
}^{\infty}} \nc{\Pzx}[2]{\prod_{#1 =0}^{#2}} \nc{\Pyx}[3]{\prod_{
#1 = #2 }^{#3}} \nc{\Izi}[2]{\int_{ 0 }^{\infty}{#1}{\rm d}\,{#2}}
\nc{\Izx}[3]{\int_{ 0 }^{#1}{#2}{\rm d}\,{#3}}
\nc{\iic}[2]{\iint_{\BC}{#1}{\rm d}^2{#2}} \nc{\qIzx}[3]{\int_{ 0
}^{#1}{#2}{\wh{\rm d}}_q\,{#3}}
\nc{\qIxy}[4]{\int_{#1}^{#2}{#3}{\wh{\rm d}}_q\,{#4}}
\nc{\qqIxy}[4]{\int_{#1}^{#2}{#3}{\text{d}}_q\,{#4}}
\nc{\qIzi}[2]{\int_{ 0 }^{\infty}{#1}{\wh{\rm d}}_q\,{#2}}
\nc{\qIii}[2]{\int_{-\infty}^{+\infty}{#1}{\wh{\rm d}}_q\,{#2}}
\nc{\qIij}[2]{\int_{-\infty}^{+\infty}{#1}{\text{d}}_q\,{#2}}
\nc{\derx}[1]{\frac{\text{d}}{\text{d}x}{#1}}
\def\qdd{\left(\ts\frac{\text{d}}{\text{d}x}\right)_q\,}
\def\qd{{\wh{\left[\ts\frac{\text{d}}{\text{d}x}\right]}}_q\,}
\def\qdy{{\wh{\left[\ts\frac{\text{d}}{\text{d}y}\right]}}_q\,}
\nc{\bhgs}[5]{{{}_{#1}{\phi}_{#2}}%
 \left(\ts{\genfrac{}{}{0pt}{}{#3}{#4}%
 \biggl.\biggr| #5 }\right)}
\nc{\sfito}[4]{{{}_2{\phi}_1}%
 \left(\ts{\genfrac{}{}{0pt}{}{#1,\,\, #2}{\phantom{\bigr(}#3}%
 \biggl.\biggr| #4 }\right)}
\nc{\sfitz}[3]{{{}_2{\phi}_0}%
 \left(\ts{\genfrac{}{}{0pt}{}{#1,\,\, #2}%
 {\phantom{\bigr(}\mbox{\rule[3pt]{7pt}{1pt}\quad}}%
 \biggl.\biggr| #3 }\right)}
\nc{\sfio}[2]{ { {}_0{\phi}_1 }%
 \left(\ts{\genfrac{}{}{0pt}{}{\mbox{\rule[3pt]{7pt}{1pt}}}{#1}%
 \biggl.\biggr| #2 }\right)}
\nc{\sfioo}[3]{{{}_1{\phi}_1}%
 \left(\!\!\ts{\genfrac{}{}{0pt}{}{#1}{\phantom{\bigr(}#2\,\,}%
 \biggl.\biggr| #3 }\right)}
\nc{\sfizo}[2]{{{}_0{\phi}_1}%
 \left(\ts{\genfrac{}{}{0pt}{}{\phantom{\bigr(}\mbox{\rule[3pt]{7pt}{1pt}
 \quad}}{#1}\biggl.\biggr| #2 }\right)}%
\def\da{a^{\dagger}}
\nc{\modx}[1]{{\vert{#1}\vert}}                 
\begin{document}
\thispagestyle{empty}

  \centerline{\Large\bf Generalized Coherent States}
  \bigskip

  \centerline{\Large\bf for q-oscillator connected with}
  \bigskip

  \centerline{\Large\bf discrete q-Hermite Polynomials
  \footnote{This research
  was supported by RFFI grants No 03-01-00837, 03-01-00593.
\hfill\break Russian version
of this article was published in {\it ZNS PDMI} {\bf 308} 48-66 (2004)}}

  \bigskip
  \centerline{\large\bf Vadim V. Borzov${}^{(1)}$
  \normalsize{and}
  \large\bf Eugene V. Damaskinsky${}^{(2)}$}

  \begin{center}
  ${}^{(1)}$
  {Department of Mathematics, St.Petersburg University of Telecommunications,}

  {191065, Moika  61, St.Petersburg, Russia}

  {vadim@VB6384.spb.edu}

\medskip

  ${}^{(2)}$
  {Department of Mathematics, University of Defense Technical Engineering,}

  {Zacharievskaya 22, St.Petersburg, Russia}

  {evd@pdmi.ras.ru}
\end{center}
\bigskip
\bigskip

\begin{quote}
{\small We are continuing here the study of generalized coherent
states of Barut-Girardello type for the oscillator-like systems
connected with the given set of orthogonal polynomials. In this
work we construct the family of coherent states associated with
discrete $q$-Hermite polynomials of the II-type and prove the
over-completeness of this family of states by constructing the
measure for unity decomposition for  this family of coherent
states. }
\end{quote}

\section {Introduction}

It is difficult to overestimate the role played by coherent states
in the quantum theory. This explains a great many of works
 in which discuss the properties of coherent
 states and their different generalizations as well as  their
numerous applications in fields of modern physics, first of all in
quantum optics. Basic facts relating to the theory of coherent
states can be found in works ~\cite{PerelomovBk}---\cite{AAZM}. An
extensive bibliography (up to early 21st century) was given in the
review~\cite{Dodo}. Only in the last few months there appeared a
number of  interesting works devoted to generalization of vector
coherent states for the case of  matrix index (~\cite
{TH46}---\cite {AEG42}), hypergeometric~\cite{AS13} and
combinatorial~\cite {BPS33} coherent states,as well as to further
development of mathematical~\cite{ART147, DK37} and physical~\cite
{GP19, A25} applications of coherent states, including
applications to the description of models with exactly solvable
potentials~\cite {SPB38}, super-symmetric conformal field
theory~\cite {Z24} and to path integral~\cite{HK9184} in
holomorphic representation.

Coherent states were first introduced for the case of the boson-
oscillator (connected with Heisenberg group)~\cite{Schr} and then
they were rediscovered in~\cite{G1}---\cite{S1}
 since quantum optics inception.Now coherent states
  are determined for a wide
class of quantum systems (including quantum field ones), and also
for systems connected with  other groups (including super-groups).
Coherent states can be determined also for quantum groups as well
as using of various exponential generalizations (deformations).

Nowadays some variant definition of coherent states are known:

1) As
eigenfunctions of the annihilation operator: $ \! a\ket {z} \! =
\! z\ket {z}, $ $z\in {\mathbb C} $ (in this case coherent states
are usually called  { \it Barut - Girardello coherent states}, so
in the work~\cite {BarutGirardello} this definition was extended
to the case of non-compact groups);

2) As a result of action of the unitary shift operator $D(z)=
e^{z\da-z^*a}$ upon the fixed vector of the state space (usually
the Fock vacuum state $\ket{0}$): $\ket{z}=D(z)\ket{0}$ ({\it
coherent states of Perelomov type}, who actively studied these
states in the research summarized in~\cite{PerelomovBk});

3) As states minimizing the Heisenberg (or Heisenberg - Robertson)
uncertainty relations;

4) As states  satisfying several natural conditions -
norm-ability, continuity for an index, (over)completeness and
 the existence of the decomposition of unity connected with it, and
sometimes evolutionary stability (the {\it coherent states of
Klauder - Gazeau type}\cite{GK,AGMKP}).

In the case of  boson- oscillator all these definitions generate
the same family of coherent states; however, it is not in a
general case.

It is known that standard coherent states of one boson-oscillator
are defined by the relation
\begin{equation}\label{bCS}
\ket{z}= e^{-\frac{\modx{z}^2}2}\; \Szi{k} \frac
{z^k}{\sqrt{k!}}\ket{k}\;,\qquad z\in\BC .
\end{equation}

The generalizations of coherent states connected with this
definition (they are known as nonlinear, generalized or deformed
coherent states) have the form
\begin{equation}\label{gCS}
\ket{z}=(\cN(\modx{z}^2))^{-\frac12}\; \Szi{k} \frac
{z^k}{\sqrt{\rho_k!}}e_k\;,\qquad z\in\cD\subseteq\BC .
\end{equation}

The  $\cN$ factor is a normalizing coefficient,
$\{e_k\}_{k=0}^{\infty}$ is orthonormal basis in Hilbert space
$\hf$ (considered also as the Fock basis and the Fock space,
respectively), $\rho_k!$ is a generalized factorial on an index
$\rho_k!=\rho_1\cdot\rho_2\cdot\ldots\cdot\rho_k$, with the
$\rho_0!=1$ constraint. The possible choice of the
 $\{\rho_k\}_{k=0}^{\infty}$ sequence of positive numbers is
limited by the (over)completeness requirement for the family of
coherent states
\begin{equation}\label{DoU}
\int_{\cD}d\mu(z,\overline{z})\; {\cN}(\modx{z}^2) \;
\ket{z}\bra{z} = I\; ,
\end{equation}
where $I$ is the identity operator in $\hf$, and the  $d\mu$
measure gives a solution of the moment problem~\cite {CMP}
connected with the $\{\rho_k\}_{k=0}^{\infty}$ sequence  (see also
\cite{bdk,sim97}). Distinct families of  generalized coherent
states differ by a distinct choice of these sequences. It is known
that if for the chosen sequence (or equivalently for the chosen
family of the generalized coherent states) the series $$ \Szi{k}
\frac{1}{\sqrt{\rho_k}}, $$ is divergent, then one can relate with
this sequence the appropriate family $\{p_k\}_{k=0}^{\infty}$ of
polynomials orthonormal with respect to the uniquely determined
measure $\text{d}\nu$ on the real line, which make up Fock basis
$\{e_k\}_{k=0}^{\infty}$ in  Fock space
$\hf=L^2({\BR};\text{d}\nu)$. These polynomials generate an
oscillator-like system, for which they play the same role as
Hermite polynomials in the case of standard boson-oscillator. The
spectrum of the hamiltonian for this system is determined by the
 $\{\rho_k\}_{k=0}^{\infty}$ coefficients of the recurrent
relations for this family of orthogonal polynomials. Note that a
specific choice of the family of the polynomials orthogonal on the
real axis connected with  Jacobi matrix (i.e. polynomials defined
by three-term recurrent relations) as the Fock basis, extends
significantly the list of applied problems of quantum optics,
where coherent states are successfully
used~\cite{BZP,NFF,KSu:bk,Dodo,OHT}.Application of generating
functions for known polynomials allows to express coherent states
 through standard special functions explicitly and to prove the completeness
of related system of coherent states by solving the appropriate
classical moment problem. It is possible also, that interpretation
of some of  differential and difference the equations arising in
this approach as Schr\"odinger equations for a system of
"generalized oscillators", will indicate the way to the solution
of certain asymptotic and spectral problems,which are well studied
for the usual Schr\"odinger equation.

Such motivation underlies the new approach to construction of
generalized coherent states, developed in the works of the authors
\cite{bd2}-\cite{bd7} (see also work~\cite{OHT}, in which the
connection of boson-systems with the families of orthogonal
polynomials is also noted). The new approach to construction of
coherent states rests on   the construction of generalized
oscillator algebras related with an arbitrary systems of
orthogonal polynomials suggested in the work~\cite{borz}. Namely,
given a system of orthogonal polynomials one can define in the
canonical way the oscillator-like system; that is the coordinate
and momentum operators, the ladder operators of creation and
annihilation satisfying the commutation relations of the deformed
boson-oscillator algebras and quadratic hamiltonian are defined.
Note that the spectrum of the latter is defined by the
coefficients of recurrent relations for these orthogonal
polynomials. This allows to construct the family of coherent
states not only for the case of classical orthogonal polynomials,
but  for their various $q$-analogues as well.

In particular situations the basic difficulties are connected with
the solution of the proper classical moment problem and with
expression of coherent states in terms of the standard special
functions. The construction of coherent states families  described
above is realized in the works of the authors~\cite{bd2} -
\cite{bd6} for the coherent states of the Barut - Girardello and
Klauder - Gazeau types, connected with the classical Hermite,
Laguerre, Legendre and Chebyshev polynomials. The main point in
this research, as well as  in the given work, is the solution of
the proper classical moment problem~\cite {CMP} (see also
\cite{bdk}), necessary for construction of the measure involved in
the decomposition of unity relation - one of the most important
properties of coherent states. The approach developed for the case
of classical polynomials was extended in \cite{bd7} to the case of
continuous $q$-Hermite polynomials. In \cite{bd7} it was shown
also, that for discrete $q$-Hermite polynomials of the I-type
coherent states do not exist. In the given work we shall continue
the investigation of the case of deformed polynomials and
construct a family of coherent states connected with the last
remaining case of discrete $q$-Hermite polynomials, namely with
the $q$-Hermite polynomials of II-type.

\nopagebreak
\section {q-oscillator, connected with discrete
$q$-Hermite polynomials of II-type}

\subsection {Discrete $q$-Hermite polynomials of II-type
$\mathbf{{\tilde h}_n(x;q)}$}

Discrete q-polynomials Hermite of II-type ${\tilde h}_n(x;q)\quad
(|q|<1)$ has the form~\cite{KS,GR}
\begin{align}
{\tilde h}_n(x;q)=i^{-n}V_n^{-1}(ix;q)&=
i^{-n}q^{-\binom{n}{2}}\sfitz{q^{-n}}{ix}{q;\,-q^n}=\nonumber\\
{}&=x^n\sfito{q^{-n}}{q^{-n+1}}{0}{q^2;\,-\frac{1}{x^2}}\,,
\end{align}
where the basic hypergeometric series ${{}_{r}{\phi}_{s}}$ by
definition is equal to
\begin{equation}
\bhgs{r}{s}{a_1,\ldots,a_r}{b_1,\ldots,b_s}{q;\,z}=
\Szi{k}(-1)^{k(1+s-r)} q^{(1+s-r)\binom{k}{2}} \frac{(a_1;q)_k
(a_2;q)_k\cdots (a_r;q)_k} {(b_1;q)_k (b_2;q)_k\cdots (b_s;q)_k}\,
\frac{z^k}{(q;q)_k}\,,
\end{equation}
so that
\begin{align}
\sfitz{a}{b}{q;\,z}&=\Szi{k}(-1)^kq^{-\binom{k}{2}}
\frac{(a;q)_k\,(b;q)_k}{(q;q)_k}\,z^k\,,
\\
\sfito{a}{b}{c}{q;\,z}&=\Szi{k}
\frac{(a;q)_k\,(b;q)_k}{(c;q)_k(q;q)_k}\,z^k\,.
\end{align}
Here the shifted q-factorials (q-Pochhammer symbols), defined by
\begin{equation}
(a;q)_0=1,\quad (a;q)_k=\Pzx{s}{k-1}(1-aq^s),\quad
(a;q)_{\infty}=\Pzi{s}(1-aq^s)
\end{equation}
are used.

Recurrent relations for polynomials ${\tilde h}_n(x;q)$ have the
form
\begin{gather}
x{\tilde h}_n(x;q)={\tilde h}_{n+1}(x;q)+ q^{-2n+1}(1-q^n){\tilde
h}_{n-1}(x;q) \label{rr1}\\ {\tilde h}_0(x;q)=1,\qquad {\tilde
h}_{-1}(x;q)\equiv 0.\nonumber
\end{gather}

The polynomial system $\left\{ {\tilde h}_n(x;q)
\right\}_{n=0}^{\infty}$ on an interval $(-\infty,+\infty)$ $
(-\infty, + \infty) $ satisfies the orthogonality relations
($m\neq n,\quad c>0$)
\begin{equation}
c(1-q)\Sii{k}\left[ {\tilde h}_{m}(cq^k;q){\tilde h}_{n}(cq^k;q)+
{\tilde h}_{m}(-cq^k;q){\tilde h}_{n}(-cq^k;q)
\right]\,W(cq^k)\,q^k=0\,,
\end{equation}
where
\begin{equation}
W(x)=\frac{1}{(ix;q)_{\infty}(-ix;q)_{\infty}}\,.
\end{equation}
One can write these orthogonality relations in the term of the
Jackson $q$-integral (see, for example, \cite{GR})
\begin{equation}
\qIij{{\tilde h}_{m}(t;q){\tilde h}_{n}(t;q)W(t)}{t}=0.
\end{equation}

Let us remind, that the concept of the Jackson $q$-integral is
 connected naturally with definition of a $q$-derivative
\begin{equation}\label{di1}
f(x):=\qdd F(x)=\frac{F(x)-F(qx)}{x(1-q)}\,,\qquad (0<q<1).
\end{equation}
From this relation it follows that
\begin{equation}\label{di2}
F(x)-F(qx)=x(1-q)f(x) \,.
\end{equation}
According to this relation, continuous in a point $x=0$ function
$F(x)$ can be reconstructed from $f(x)$ by  the formula
\begin{align}
F(x)-F(0)&=\left( F(x)-F(qx) \right) + \left( F(qx)-F(q^2x)
\right) +\cdots+\left( F(q^nx)-F(q^{n+1}x) \right)
+\cdots\nonumber\\ {}&=x(1-q)\Szi{n}q^nf(q^nx)\,. \label{di3}
\end{align}
From \reff{di3}, using an analogue of the Newton - Leibnitz
formula, we obtain
\begin{equation}\label{di4}
\qqIxy{0}{x}{f(t)}{t}:=F(x)-F(0)=x(1-q)\Szi{n}q^nf(q^nx)\,.
\end{equation}
Similarly, we receive
\begin{equation}\label{di5}
\qqIxy{x}{\infty}{f(t)}{t}:=x(1-q)\Syx{n}{-\infty}{-1}q^nf(q^nx)\,.
\end{equation}
Adding \reff{di4} and \reff{di5}, we receive for all $x>0$
\begin{equation}\label{di6}
\qqIxy{0}{\infty}{f(t)}{t}:=x(1-q)\Syx{n}{-\infty}{\infty}q^nf(q^nx)\,.
\end{equation}
In the same way
one can to define
\begin{equation}\label{di7}
\qqIxy{-\infty}{0}{f(t)}{t}:=x(1-q)\Syx{n}{-\infty}{\infty}q^nf(-q^nx)\,.
\end{equation}
Then for all $x>0$ we have
\begin{equation}\label{di7a}
\qqIxy{-\infty}{\infty}{f(t)}{t}:=x(1-q)\Syx{n}{-\infty}{\infty}q^n
\left[f(q^nx)+f(-q^nx)\right]\,.
\end{equation}
Let us note that for simplification of notation, in these formulas
one  choose often $x=1$.

 For the case of the "symmetric" $q$-derivative
($\qdd F(x):=\frac{F(qx)-F(q^{-1}x)}{x(q-q^{-1})}$) the definition
of the Jackson $q$-integral was given in~\cite{DK189}). Below we
construct  an other variant of the Jackson $q$-integral, connected
with the deformed derivative $\qd$.

Following the construction given in \cite{borz}, we introduce
normalized family of polynomials
 $\left\{\Psi_n(x;q)\right\}_{n=0}^{\infty}$,
by the relation
\begin{equation}\label{rr2}
\Psi_{n}(x;q)=\frac{q^{\half n^2}}{\sqrt{(q;q)_{n}}} \,{\tilde
h}_{n}(x;q)\,,
\end{equation}
 From \reff{rr1} we obtain a recurrent relation
\begin{align}
\Psi_{n}(x;q)=b_{n}\Psi_{n+1}(x;q)+b_{n-1}\Psi_{n-1}(x;q)\,,\label{rr2}
\\ \intertext{where} b_{n}=q^{-\half
(2n+1)}\sqrt{1-q^{n+1}\,}\,,\quad b_{-1}\equiv 0\,.
\end{align}

The constructed family of polynomials forms orthonormal basis in
the space $\chm={\rm L}^2(\BR;{\rm d}\mu)$ with respect to some
probability measure  ${\rm d}\mu(x).$ Because of
\begin{equation}
\Szi{n}b_{n}^{\,-1}<\infty\,,
\end{equation}
the moment problem for the Jacobi matrix, defined by the recurrent
relations \reff{rr2}, is undetermined (see \cite{CMP}) and the
measure ${\rm d}\mu(x)$ is determined not uniquely.

In the following we shall consider the Hilbert space $\chm$ as the
Fock space for oscillator - like system (q-oscillator) which will
be constructed in the following subsection.

\subsection {q-oscillator, connected with discrete
 $q$-Hermite polynomials of II-type}
Let us define the position operator $X_{\mu}$ in the Hilbert space
$\chm$ by its action
\begin{equation}
X_{\mu}\ket{n}=X_{\mu}\Psi_{n}(x;q)=b_{n}\ket{n+1}+b_{n-1}\ket{n-1},
\label{Xop}
\end{equation}
on the basis elements $\ket{n}:=\Psi_{n}(x;q),$ $n=0,1,2,\ldots$,
in full accordance with the recurrent relations \reff{rr2}. It is
known (see \cite{CMP}), that such defined operator $X_{\mu}$ is
symmetric and has indices of defect equal to $(1,1),$ so it has
the family of the selfadjoint extensions. Following the method
described in \cite{bdk}, fixing some selfadjoint extension
$X_{\mu}^{\vp}$ of the operator $X_{\mu}$, it is possible to
construct the completely defined extremal measure ${\rm
d}\mu_{\vp}$. Let us recall some related results from \cite {bdk}
for a case when parameter $\vp$ numbering extensions has the zero
value $\vp=0.$ In this case the carrier $\Pi_0\equiv
\left\{x_k\right\}_{k=-\infty}^{\infty}$ of the measure ${\rm
d}\mu_{0}$ consists from the points being the roots of
transcendental equation
\begin{align}
\Psi_{0}(x;q)+x\Szi{k} (-1)^k\sqrt{\frac{[2k-2]!!}{[2k-1]!!}\,}\,
\Psi_{2k-1}(x;q)=0\,, \\ \intertext{where} [s]\equiv
\frac{b_{s-1}^2}{b_{0}^2}\,,\quad s\geq1\,,\qquad [0]=0\,.
\end{align}
Loadings $\sigma_0(x_k)$ in a points of the carrier $\Pi_0\equiv
\left\{x_k\right\}_{k=-\infty}^{\infty}$ of the measure ${\rm
d}\mu_{0}$ are equal
\begin{equation}
\sigma_0(x_k)=\frac{x_k\Syx{j}{1}{\infty}\frac{1}{[2j-1]!!}
\Syx{m}{0}{j-1}(-1)^{j+m+1}\beta_{2m,\,2j-2}{x_k}^{2(j-m-1)}}
{\left(\derx{\left[ -1+\Syx{j}{1}{\infty}\frac{1}{[2j-1]!!}
\Syx{m}{0}{j-1}(-1)^{j+m-1}\alpha_{2m-1,\,2j-2} x^{2(j-m)}
\right]}\right)(x_k)}.
\end{equation}
Coefficients $\alpha_{i\,j}$ and $\beta_{i\,j}$ are coefficients
of polynomials $P_n(x;q)=\Psi_n(x;q)$ and $Q_n(x;q)$ of the 1-st
and 2-nd type for the Jacobi matrix connected with the operator
$X_{\mu}$ \cite{CMP}. We recall their explicit expressions from
the work~\cite{bdk}
\begin{align}
P_n(x;q)&=\Szx{m}{\text{Ent}(\half n)}\frac{(-1)^m}{\sqrt{[n]!\,}}
\alpha_{2m-1,\,n-1}x^{n-2m}\,,\quad n\geq0\,, \\
\alpha_{-1,\,2n-1}&=1,\qquad \alpha_{2m-1,\,n-1}=
\Syx{k_1}{2m-1}{n-1}[k_1]\Syx{k_2}{2m-3}{k_1-2}[k_2]\ldots
\Syx{k_m}{1}{k_{m-1}-2}[k_m]\,,\quad m\geq1\,.
\end{align}
\begin{align}
Q_0(x;q)&=0\,,\qquad Q_{n+1}(x;q)= \Szx{m}{\text{Ent}(\half
n)}\frac{(-1)^m}{\sqrt{[n+1]!\,}} \beta_{2m,\,n}\,x^{n-2m}\,,\quad
n\geq0\,, \\ \beta_{0,\,n}&=1,\qquad \beta_{2m,\,n}=
\Syx{k_1}{2m}{n}[k_1]\Syx{k_2}{2m-2}{k_1-2}[k_2]\ldots
\Syx{k_m}{2}{k_{m-1}-2}[k_m]\,,\quad m\geq1\,.
\end{align}

Let us note, that for some problems there is no necessity to
consider an extremal measure. In that case it is possible to take
the most "convenient" carrier of the measure, for example, a
geometrical progression $\left\{cq^k\right\}_{k=-\infty}^{\infty}$
and pick up loadings so that to receive the solution of the same
moment problem (see detailed consideration in \cite{Berg1}).

We define the momentum operator $P_{\mu}$ by its action on the
basic elements of the space $\chm$
\begin{equation}
P_{\mu}\Psi_{n}(x;q)=i\left( b_{n}\Psi_{n+1}(x;q)-
b_{n-1}\Psi_{n-1}(x;q)\right)\,,
\end{equation}
or in other notation
\begin{equation}
P_{\mu}\ket{n}=i\left( b_{n}\ket{n+1}-b_{n-1}\ket{n-1}\right)\,.
\end{equation}
This operator as well as $X_{\mu}$ is symmetric and also has
defect indexes $(1,1),$ so it too has a family of self-adjoint
extensions. We shall consider that self-adjoint extension of the
operator $P_{\mu}$ which is correspond to the self-adjoint
extension of the operator $X _ {\mu} $ chosen above so, that the
related creation and annihilation operators are adjointed. Let us
keep for these extensions the same notation $X_{\mu}$ and
$P_{\mu}.$ Following the methods described in~\cite{bd1}, one can
realize the operator $P_{\mu}$ in space $\chm$ as differential -
difference operator.

Let us define the ladder operators of creation and annihilation in
the space $\chm$ by the relations
\begin{equation}
\apm=\half\sqrt{\frac{q}{1-q}\,}\left(X_{\mu}-iP_{\mu}\right)\,,
\qquad
\amm=\half\sqrt{\frac{q}{1-q}\,}\left(X_{\mu}+iP_{\mu}\right)\,.
\end{equation}
These operators act on the basic elements in the space $\chm$,
according to relations
\begin{gather}
\apm\Psi_{n}(x;q)=\sqrt{\frac{q}{1-q}\,}b_{n}\Psi_{n+1}(x;q),\quad
n\geq0\,;\\
\amm\Psi_{n}(x;q)=\sqrt{\frac{q}{1-q}\,}b_{n-1}\Psi_{n-1}(x;q),\quad
n\geq1\,; \qquad\amm\Psi_{0}(x;q)=0\,.
\end{gather}
Let us define further the operator $N$, numbering the basic
states,
\begin{equation}
N\Psi_{n}(x;q)=n\Psi_{n}(x;q)\,;\quad
\left(N\ket{n}=n\ket{n}\right)\, \quad n\geq0\,.
\end{equation}
It is easy to check the validity of the following relations
\begin{equation}\label{aa}
\amm\apm=q^{-2N}\qN{N+I}\,,\qquad \apm\amm=q^{-2(N-1)}\qN{N}\,,
\end{equation}
where a symbol $\qN{n}:=\ds\frac{1-q^n}{1-q}$ denotes the standard
"mathematical" q-number. From formulas \reff{aa} one obtains the
following commutation relations for the creation and annihilation
operators:
\begin{equation}
\amm\apm -q^{-1}\apm\amm=q^{-2N}\,,\qquad \amm\apm
-q^{-2}\apm\amm=q^{-N}\,.
\end{equation}

The polynomials  $\ket{n}:=\Psi_{n}(x;q),$  are eigenfunctions
\begin{equation}\label{ev}
H_{\mu}\ket{n}=\lambda_n\ket{n}\,,
\end{equation}
of the hamiltonian
\begin{equation}
H_{\mu}=\half\frac{q}{1-q}\left({X_{\mu}}^2+{P_{\mu}}^2\right)=
\apm\amm+\amm\apm
\end{equation}
corresponding to eigenvalues
\begin{equation}
\lambda_n=\frac{q}{1-q}\left( b_{n-1}^{\,2}+b_{n}^{\,2}\right)
=q^{-2n}\qN{n+1}+q^{-2(n-1)}\qN{n}\,,\quad n\geq0\,.
\end{equation}

In the same way as in \cite{borz} where the case of the classical
polynomials is considered, it is possible to prove, that the
equation \reff{ev} is equivalent to the q-difference equation for
discrete q-Hermite polynomials of the II-nd type:
\begin{equation}
-(1-q^n) x^2 {\tilde h}_n(x;q) =q {\tilde h}_n(x-i;q) -
(1+q+x^2){\tilde h}_n(x;q) + (1+x^2) {\tilde h}_n(x+i;q)\,,
\end{equation}
which is analogue of the Schr\"odinger equation for the
$q$-oscillator described above.

\section {coherent states for q-oscillator, associated with
discrete q-Hermite polynomials of the II-nd type ${\tilde
h}_n(x;q)$}

Let us define the Barut - Girardello coherent states by the
standard way~\cite{KPS}
\begin{equation}\label{cs}
\amm\ket{z}=z\ket{z}\,;\qquad
\ket{z}={\cN}^{-1}(|z|^2)\,\Szi{n}\frac{z^n}
{\left(\sqrt{\frac{q}{1-q}\,}b_{n-1}\right)!}\,\ket{n}\,.
\end{equation}
Taking into account, that
\begin{equation}
\left( \frac{q}{1-q} {b_{n-1}}^2 \right)!=
\left(\frac{q}{1-q}\right)^n\,q^{-n^2}(q;q)_n
\end{equation}
one can calculate a normalizing factor
\begin{equation}\label{cN2}
{\cN}^{2}(|z|^2)=\Szi{n}\left(\frac{1-q}{q}\,|z|^2\right)^n\,
\frac{q^{n^2}}{(q;q)_n}=\sfio{0}{q;(1-q)|z|^2}\,.
\end{equation}
and overlapping of two coherent states
\begin{equation}
\pair{z_1}{z_2}=\Szi{n}\left(\frac{1-q}{q}\,\bar{z}_1z_2\right)^n\,
\frac{q^{n^2}}{(q;q)_n}=\sfio{0}{q;(1-q)\bar{z}_1z_2}\,.
\end{equation}

Further, substituting \reff{cN2} in \reff{cs} and using
reproducing function for polynomials ${\tilde h}_n(x;q)$  (see
\cite{KS} (3.296)):
\begin{equation}
\Szi{n}{\tilde h}_n(x;q)\tau^n=
(i\tau;q)_{\infty}\sfioo{ix}{i\tau}{q;\,-i\tau},
\end{equation}
we receive ,taking into account \reff{rr2}, the following explicit
expression for the coherent state $\ket{z}$
\begin{align}
\ket{z}&={\cN}^{-1}(|z|^2)\,\Szi{n}
\frac{z^n}{\sqrt{\left(\frac{q}{1-q}\right)^n\,q^{-n^2}(q;q)_n\,}}\,
\frac{{\tilde h}_n(x;q)}{q^{-\half
n^2}\sqrt{(q;q)_n\,}}=\nonumber\\
{}&=\left(\sfio{0}{-iq;(1-q)|z|^2}\right)^{-\half}\,\Szi{n}
\left(\sqrt{\frac{1-q}{q}\,zq}\right)^n\,\frac{q^{n(n-1)}}{(q;q)_n}\,
{\tilde h}_n(x;q)=\nonumber\\
{}&=\frac{(i\sqrt{q(1-q)\,}z;q)_{\infty}
\,\sfioo{ix}{i\sqrt{q(1-q)\,}z}{q;\,-i\sqrt{q(1-q)\,}z}}
{\sqrt{\sfio{0}{-iq;(1-q)|z|^2\,}}}\,.
\end{align}

It is necessary to prove the decomposition of unity formula (a
(over)completeness relation)
\begin {equation} \label {ocr1}
\iic {\wh {W} (|z | ^ 2) \ket {z} \bra {z}} {z} = \one \,
\end {equation}
i.e. to construct a measure
\begin{equation}\label{ocr2}
{\rm d}\mu(|z|^2)=\wh{W}(|z|^2){\rm d}^2z\,.
\end{equation}

It is known \cite{KPS,PS} that for this purpose it is necessary to
solve a classical Stieltjes moment problem
\begin{gather}
\pi\Izi{x^nW(x)}{x}=\left(\frac{q}{1-q}\right)^n\,q^{-n^2}(q;q)_n\,,
\quad n\geq0\,,\label{m1}\\ \intertext{Ј¤Ґ}
W(x)=\frac{\wh{W}(x)}{{\cN}^{2}(x)},\quad (x=|z|^2).
\end{gather}
Under the replacement of variables
\begin{equation}
x=\frac{q}{1-q}y\,,\qquad \frac{\pi
q}{1-q}W(\frac{q}{1-q}y)=\wt{W}(y)\,,
\end{equation}
the moment problem \reff{m1} takes the form
\begin{equation}\label{m2}
\Izi{y^n\wt{W}(y)}{y}=q^{-n^2}(q;q)_n\,.
\end{equation}

\section {The solution of the moment problem {\protect(\ref{m2})}}

Let us define generalized exponential
\begin{equation}
\gex{x}:=\Szi{n}\frac{x^n}{b_{n-1}!}=\Szi{n}\frac{q^{n^2}}{(q;q)_n}x^n=
\sfio{0}{q,\,qx}
\end{equation}
and the corresponding deformed derivative
\begin{equation}\label{qd}
\qd f(x)=\frac{f(q^{-2}x)-f(q^{-1}x)}{q^{-1}x},\
\end{equation}
so that
\begin{equation}
\qd x^n=\left\{\begin{aligned} {b_{n-1}}^2
x^{n-1}&\quad\text{ЇаЁ}\quad n\geq1;\\
0\qquad&\quad\text{ЇаЁ}\quad n=0
\end{aligned}\right.
\,,\qquad\qquad \qd\gex{x}=\gex{x}\,;
\end{equation}
and following generalization of Leibnitz rule
\begin{equation}\label{LR}
\qd\left(u(x)v(x)\right)= \left\{\begin{aligned} v(q^{-2}x)\qd
u(x)&+u(q^{-1}x)\qd v(x)\,,\\ v(q^{-1}x)\qd u(x)&+u(q^{-2}x)\qd
v(x)\,.
\end{aligned}\right.
\end{equation}
is hold.

The related q-integral is defined by  equality
\begin{equation}\label{qi1}
\qIzi{f(t)}{t}=\frac{1-q}{q^2}\Szi{k}
\left(q^{-(k-1)}f(q^{-(k-1)})+q^{(k+2)}f(q^{(k+2)})\right)\,,
\quad |q|<1\,.
\end{equation}

It is not difficult to prove that this integral exists for
continuous functions
$f(x)=\qd F(x),$  such that
$F(x)\stackrel{x\rightarrow\infty}{\longrightarrow}0$ and that
\begin{equation}\label{qi2}
\qIzx{a}{\qd f(x)}{x}=f(a)-f(0)\,.
\end{equation}
after integrating the Leibnitz rule \reff{LR} in view of the
relation \reff {qi2}, we receive the formula of integration by
parts ($0<a\leq\infty$)
\begin{subequations}   \begin{eqnarray}
\qIzx{a}{u(q^{-1}x)\qd v(x)}{x}
  =u(x)v(x)\,\rule[-6pt]{.5pt}{18pt}^{\,a}_{\,0}-
  \qIzx{a}{v(q^{-2}x)\qd u(x)}{x}\,;\label{ip1}
\\
\qIzx{a}{u(q^{-2}x)\qd v(x)}{x}
  =u(x)v(x)\,\rule[-6pt]{.5pt}{18pt}^{\,a}_{\,0}-
  \qIzx{a}{v(q^{-1}x)\qd u(x)}{x}\,.\label{ip2}
\end{eqnarray} \end{subequations}

For a case $a=\infty$ we shall rewrite the relation \reff{ip1} in
the following form. At first, having done the replacement
$q^{-1}x=y$, in the integral in the right hand part of equality
\reff{ip1} we shall receive
\begin{equation}
\qIzi{u(q^{-1}x)\qd v(x)}{x}=\qIzi{u(y)\qd v(qy)}{y}\,.
\end{equation}
 Taking into account independence of integral \reff{qi1} from
a designation of the variable of integration, we shall receive
finally
\begin{equation}
\qIzi{u(x)\qd v(qx)}{x}
  =u(x)v(x)\,\rule[-6pt]{.5pt}{18pt}^{\,\infty}_{\,0}-
  \qIzi{v(q^{-2}x)\qd u(x)}{x}\,.\label{ip3}
\end{equation}

Let us consider the function $f(y)$, satisfying a condition
\begin{equation}\label{312}
\qdy f(q^2y)=-q^{2}f(y)\,.
\end{equation}
Using the definition of the deformed derivative \reff{qd}, we
receive from \reff{312} the difference equation $$
\frac{f(qy)-f(q^2y)}{q^{-1}y}=-q^{2}f(y) $$ for the function
$f(y).$ Rewriting this equation in a form
\begin{equation}\label{313}
f(qy)-f(q^2y)=-qyf(y)\,,
\end{equation}
we find his solution in the form of power series
\begin{equation}\label{314}
f(y)=\Szi{n}\frac{1}{(q;q)_n}q^{-\binom{n}{2}} (-y)^n=
\sfitz{0}{0}{q;\,y}\,.
\end{equation}

Let us consider an integral
\begin{equation}\label{315}
I_n(q):=\qIzi{x^n\,\sfitz{0}{0}{q;\,q^{-2}x}}{x}\,,\quad n\geq0\,.
\end{equation}
From \reff{312} it follows, that
\begin{equation}\label{316}
-\qd\sfitz{0}{0}{q;\,qx}=\sfitz{0}{0}{q;\,q^{-2}x}\,.
\end{equation}
Applying to an integral $I_n(q)$ the formula of integration by
parts \reff{ip3} as $u(x)=x^n$ and
$v(x)=-\left(\sfitz{0}{0}{q;\,x}\right),$ we obtain for $n\geq1$
\begin{equation}\label{317}
I_n(q)=\qIzi{\sfitz{0}{0}{q;\,q^{-2}x}\,{b_{n-1}}^2x^{n-1}}{x}\,.
\end{equation}
Because the out of integral term vanishes, we obtain finally
\begin{equation}\label{318}
I_n(q)={b_{n-1}}^2I_{n-1}(q)\,,\quad n\geq1\,.
\end{equation}
On the other hand, for $n=0 $ we have
\begin{equation}\label{319}
I_0(q)=-\qIzi{\qd\sfitz{0}{0}{q;\,q^{-2}x}}{x}=
-\sfitz{0}{0}{q;\,qx}\,\rule[-6pt]{.5pt}{18pt}^{\,\infty}_{\,0}=1.
\end{equation}
From the relations \reff{318} and \reff{319} it follows that
\begin{equation}\label{320}
I_n(q)={b_{n-1}}^2!=q^{-n^2}(q;q)_n\,.
\end{equation}

Thus, the classical moment problem \reff{m2} has the following
solution
\begin{equation}\label{321}
\wt{W}(y)= \frac{1-q}{q^2}\Szi{k}y\,\,\sfitz{0}{0}{q;\,q^{-2}y}
\left(\delta(y-q^{-(k-1)})+\delta(y-q^{(k+2)})\right)
\end{equation}
so that
\begin{multline}\label{322}
\wh{W}(x)={\cN}^2(x)\frac{1-q}{\pi q} \wt{W}(\tfrac{1-q}{q}
x)=\frac{1-q}{\pi q}\sfizo{0}{q;\,(1-q)x}\times\\
\Szi{k}\frac{(1-q)^2}{q^3} x\,\,\sfitz{0}{0}{q;\,\tfrac{1-q}{q^3}
x} \left(\delta(\tfrac{1-q}{q} x-q^{-(k-1)}){\phantom{\biggr(}}
      +\delta(\tfrac{1-q}{q} x-q^{(k+2)})\right)\,.
\end{multline}
Thus for a measure $\text{d}\mu(|z|^2)$ from the decomposition of
unity (\ref{ocr1}) - (\ref{ocr2}) we have
\begin{multline}\label{323}
\text{d}\mu(|z|^2)= \frac{(1-q)^3}{\pi
q^4}\,\sfizo{0}{q;\,(1-q)|z|^2}\times\\
\Szi{k}|z|^2\,\,\sfitz{0}{0}{q;\,\tfrac{1-q}{q^3} |z|^2}
\left(\delta(\tfrac{1-q}{q}|z|^2-q^{-(k-1)}){\phantom{\biggr(}}
      +\delta(\tfrac{1-q}{q}|z|^2-q^{(k+2)})\right)\,.
\end{multline}
Thus completeness of the constructed system of coherent states is
proved.

\bigskip
\bigskip

{\bf Acknowledgments}. Authors are grateful to the prof.
V.M.Babich and to participants of his seminar for discussion the
results closely connected with theme of the present work. Authors
are grateful to the prof. P.P.Kulish for attention to work and
useful remarks. The given research was carried out under the
support of the RFBR grants No 03-01-00837, 03-01-00593.

\def\pra#1#2#3#4{{\it Phys. Rev. A.} {\bf #1}, {#2} {#3} {(#4);}}
\def\jmp#1#2#3#4{{\it J. Math. Phys.} {\bf #1}, {#2} {#3} {(#4);}}

\end {document}